**Epitaxial growth, structural characterization and exchange bias of non-collinear antiferromagnetic $Mn_3Ir$ thin films**

James M. Taylor * [1], Edouard Lesne [1], Anastasios Markou [2], Fasil Kidane Dejene [1], Benedikt Ernst [2], Adel Kalache [2], Kumari Gaurav Rana [1], Neeraj Kumar [1], Peter Werner [1], Claudia Felser [2], Stuart S. P. Parkin † [1]

[1] Max Planck Institute of Microstructure Physics, Weinberg 2, 06120 Halle (Saale), Germany

[2] Max Planck Institute for Chemical Physics of Solids, Nöthnitzer Str. 40, 01187 Dresden, Germany

## Abstract

Antiferromagnetic materials are of great interest for spintronics. Here we present a comprehensive study of the growth, structural characterization, and resulting magnetic properties of thin films of the non-collinear antiferromagnet $Mn_3Ir$. Using epitaxial engineering on MgO (001) and $Al_2O_3$ (0001) single crystal substrates, we control the growth of cubic γ-$Mn_3Ir$ in both (001) and (111) crystal orientations, and discuss the optimization of growth conditions to achieve high-quality crystal structures with low surface roughness. Exchange bias is studied in bilayers, with exchange bias fields as large as -29 mT (equivalent to a unidirectional anisotropy constant of 11.5 nJ $cm^{-2}$) measured in $Mn_3Ir$ (111) / permalloy heterostructures at room temperature. In addition, a distinct dependence of blocking temperature on in-plane crystallographic direction in $Mn_3Ir$ (001) / Py bilayers is observed. These findings are discussed in the context of chiral antiferromagnetic domain structures, and will inform progress towards topological antiferromagnetic spintronic devices.

* james.taylor@mpi-halle.mpg.de

† stuart.parkin@mpi-halle.mpg.de

## I - Introduction

Artificial (or synthetic) antiferromagnetic structures (SAFs) [1,2] have played a key role in spintronics since the invention of the spin-valve sensor for detecting tiny magnetic fields in magnetic recording read heads [3] and in magnetic tunnel junction (MTJ) memory bits for magnetic random access memory (MRAM) applications [4]. More recently, highly efficient current driven motion of domain walls in SAFs was discovered [5], that makes possible racetrack memory devices [6] by utilizing the chirality of the magnetic structure [7]. SAFs and related multilayers are used to eliminate long-range magnetostatic fields that otherwise make nanoscopic spin-valves and MTJs inoperable. Furthermore, the resonance frequencies of antiferromagnet (AF) materials can be much higher than ferromagnet (FM) materials [8], making such materials of interest for ultrafast spin dynamics.

Motivated by these improvements in performance, the field of antiferromagnetic spintronics has rapidly grown [9], investigating a range of different materials. More recent experimental observations include spin-orbit torque switching and electrical read-out of the AF state in CuMnAs [10], $Mn_2Au$ [11] and MnTe [12], as well as spin currents and spin Hall magnetoresistance effects in PtMn [13,14]. Understanding of AF domain structure is important for the efficient control of the above effects in these metallic materials, all of which exhibit collinear AF order.

Materials with a non-collinear spin texture, such as $Mn_3X$ ($X$ = Ir, Pt, Sn, Ge), are promising candidates for *topological* AF spintronic applications [15]. This follows the prediction of an intrinsic anomalous Hall effect (AHE) in the $L1_2$ ordered phase of cubic $Mn_3Ir$ [16]. In addition, a facet-dependent spin Hall effect (SHE) has been measured in epitaxial thin films of $Mn_3Ir$ [17], whose origin derives from a Berry curvature-driven effective field generated by a combination of spin-orbit coupling and symmetry breaking arising from the *chiral* AF structure [17,18].

For the related compounds $Mn_3Sn$ and $Mn_3Ge$, subsequent to theoretical predictions [19], large AHE has been experimentally demonstrated in highly-ordered bulk samples [20,21]. This has been enabled by the ability to align a small geometrically frustrated uncompensated in-plane magnetization via an external magnetic field, in turn coherently orienting triangular spin texture throughout the material and driving the system into a dominant chiral domain state [22]. Thus, the manipulation of AF domain structure is critical to the utilization of these phenomena.

Whilst $Mn_3Sn$ has only recently been grown in (0001) c-axis oriented epitaxial thin films [23], cubic $Mn_3Ir$ has been extensively studied in the context of exchange bias, where textured polycrystalline phases [24,25] were shown to yield the largest effects in pinning the reference magnetic electrode in spin valves and MTJs [4,26]. Later developments have proceeded to use $Mn_3Ir$ as the active element in such AF/FM heterostructures, acting as a source of spin current via SHE [27-29] and in turn generating spin-orbit torques [30] resulting in technologically attractive field-free switching of magnetic layers [31,32] desired in high-density MRAM. All of these implementations have utilized polycrystalline thin films.

On the other hand, the elucidation of novel Berry curvature-driven phenomena arising from the non-collinear spin texture requires test-bed materials with well-controlled microstructure and crystallographic properties. To this end, in this work we report detailed procedures for the preparation of epitaxial thin films of cubic $Mn_3Ir$ with both (001) and (111) orientations. Their high-quality epitaxial growth is demonstrated through a comprehensive study of crystal structure. Measurements of the exchange bias induced in bilayers with FMs is used to investigate the magnetic state of the films, which is discussed in the context of topological domain configuration. In linking our results to crystal microstructure, we underline the importance of its control when utilizing non-collinear AF films in future chiral spintronic devices.

**II – Experimental methods**

$Mn_3Ir$ thin films were deposited by magnetron sputtering in a BESTEC UHV system with base pressure < $9 \times 10^{-9}$ mbar, using a process Ar gas pressure of $3 \times 10^{-3}$ mbar. The substrate-target distance was fixed at ≈ 150 mm, while substrates were rotated to aid homogenous growth. Thin film samples were grown in both (001) and (111) crystal orientations with various thicknesses, as follows: MgO (001) [Substrate] / $Mn_3Ir$ (001) [3 or 10 nm] / TaN [2.5 nm] and $Al_2O_3$ (0001) [Substrate] / TaN (111) [5 nm] / $Mn_3Ir$ (111) [3 or 10 nm] / TaN [2.5 nm]. MgO and $Al_2O_3$ substrates were ultrasonically cleaned in acetone and ethanol, then clamped mechanically to a holder, and subsequently heated to 250 °C and 500 °C respectively under vacuum for 30 min before deposition.

$Mn_3Ir$ was grown from a $Mn_{80}Ir_{20}$ alloy target which, with DC sputtering power of 100 W, resulted in a composition of $Mn_{(0.72 \pm 0.03)}Ir_{(0.28 \pm 0.03)}$, that was determined by a combination of Rutherford backscattering spectroscopy (RBS) and energy dispersive X-ray spectroscopy (EDXS). As expected from the MnIr phase diagram, this composition allows the system to form the stoichiometric $Mn_3Ir$ phase with face-centered-cubic (fcc) crystal structure [33]. A TaN capping layer was subsequently grown in-situ from a Ta target by RF reactive sputtering at 150 W, with 33 vol.% $N_2$ partial flow introduced to the sputtering gas mixture, resulting in a composition $Ta_{(0.52 \pm 0.05)}N_{(0.48 \pm 0.05)}$ as inferred from RBS.

The TaN growth rate was 0.6 Å $s^{-1}$, while the $Mn_3Ir$ growth rate was 1.2 Å $s^{-1}$. These were measured using a quartz crystal microbalance, and deposition times adjusted to obtain desired nominal film thicknesses. Actual thicknesses were subsequently confirmed by measuring x-ray reflectivity, with fits to the data yielding individual layer thicknesses, as shown in Fig. 1(a). Fringes are observed up to high reflectivity angles, indicating the growth of smooth films with sharp interfaces. Two thicknesses of $Mn_3Ir$ (3 and 10 nm) were chosen such that their exchange bias blocking temperatures lie either below or above room temperature respectively [34].

The films' crystal structure was investigated using a combination of X-ray diffraction (XRD) and transmission electron microscopy (TEM). XRD was performed using a PANalytical X'Pert[3] diffractometer with Cu $K_{\alpha1}$ radiation ($\lambda$ = 1.5406 Å). Plane-view high-resolution TEM and cross-sectional high-angle annular dark field scanning TEM (HAADF-STEM) were measured using an FEI Titan 80-300 microscope, after fabricating thin lamella via focused ion beam milling.

**III - Film growth and structural characterization**

$Mn_3Ir$ films with a (001) orientation were achieved by growing on (001) cut single crystal MgO substrates [35]. $Mn_3Ir$ was deposited as described above, at different elevated substrate temperatures (growth temperature, $T_G$). Fig. 1(b) shows specular out-of-plane (OP) 2Θ-ω XRD patterns for the resulting 10 nm thick films. In all cases a $Mn_3Ir$ (002) diffraction peak is observed, indicating growth of $Mn_3Ir$ with a (001) crystal orientation. From the full width at half maximum (FWHM) of this peak, the size of crystalline grains in the OP direction can be estimated using the Scherrer formula, as plotted in Fig. 2(a). The intensity of the (002) peak increases with $T_G$, whilst its FWHM decreases. This indicates the (001) texture of the film strengthens with increasing $T_G$, due to the growth of larger grains of consistent crystal orientation. Such a process is further enhanced by in-situ annealing for 60 min at 500 °C.

The variation of OP lattice parameter, *c*, is also plotted in Fig. 2(a). As $T_G$ is increased, *c* relaxes towards the bulk value, presumably because of a combination of the elevated temperature improving adatom mobility and the lower thermal expansion coefficient of the insulating substrate with respect to the metallic film. This process is accentuated by post-annealing, after which *c* approaches the bulk value of 3.780 Å.

The average roughness of the $Mn_3Ir$ layers as a function of growth temperature is plotted in the inset of Fig. 1(b), as measured from atomic force microscopy (AFM) studies. The roughness increases markedly for $T_G$ above 300 °C and after post-annealing. As such, subsequent samples were grown at 300 °C without post-annealing, to achieve a compromise between high-quality crystal structure and a smooth surface. Fig. 2(b) shows an AFM topographical map from a 3 nm thick film grown under such conditions, where terraces of the MgO substrate can be seen stacked along the [100] crystal axis, with the $Mn_3Ir$ following these and showing a low root mean square (RMS) roughness of ~8 Å (measured over an area of 25 $\mu m^2$).

In this growth mode, $Mn_3Ir$ has fcc crystal lattice, and can grow in either an $L1_2$ ordered phase ($Pm\bar{3}m$, space group = 221), the crystal and magnetic structure of which is shown in Fig. 3(a), or a γ disordered phase ($Fm\bar{3}m$, space group = 225) [36]. Contrary to previous reports, no (001) superstructure peak from $Mn_3Ir$ is observed in the XRD patterns in Fig. 1(b) [37,38]. Instead, these thin films grow in the γ-$Mn_3Ir$ phase, possessing the non-collinear AF order determined by Kohn *et al.* [36], where the Mn moments have been shown to cant slightly out of the (111) plane.

To evaluate the in-plane (IP) orientation of the $Mn_3Ir$ (001) thin films grown at 300 °C, pole figure XRD measurements were performed in which the azimuthal angle Φ is scanned as a function of tilt angle χ, with 2Θ-ω fixed at the (111) reflection of a 10 nm $Mn_3Ir$ (001) film. The [100] and [010] edges of the MgO substrate were aligned along ϕ = 0 ° and ϕ = 90 ° respectively. The resulting map is shown in Fig. 3(b), with the four sharp peaks demonstrating well-defined IP crystal axes arising from a highly-oriented thin film with cubic symmetry [37,39].

Furthermore, the peak positions indicate cube-on-cube growth with respect to the MgO substrate, with the epitaxial relationship: MgO (001) [100] || $Mn_3Ir$ (001) [100]. A schematic illustration of this relationship is displayed in Fig. 3(a).

Having determined their IP orientation, 2Θ-ω XRD scans (longitudinal dimension) at different ω offset angles (transverse dimension) were recorded such that the (111) reflections from both the 10 nm $Mn_3Ir$ (001) film and the MgO (001) substrate were observed. Fig. 4(a) shows a map of such off-specular ($\Theta \neq \omega$) measurements, in which the FWHM of the $Mn_3Ir$ peak in the transverse scans along the ω axis demonstrates a mosaicity ≈ 2 °, comparable with that reported for MBE grown films [36]. Meanwhile, the FWHM in longitudinal scans along the 2Θ axis can be converted to an average crystallite size of (10.6 ± 0.7) nm, comparable with other reports [37]. This contains both IP and OP contributions to grain size suggesting that, since vertical grain size was already determined to be slightly below film thickness, crystallites grow larger laterally. This is confirmed by plane-view TEM measurements, displayed in Fig. 4(b). The in-plane microstructure of the sample is visible, with high-contrast produced between crystallites with small mosaic spread, showing lateral grain sizes between 10 and 15 nm.

Finally, the IP lattice parameter, *a*, for a 10 nm $Mn_3Ir$ (001) film grown at 300 °C was calculated to be *a* = (3.808 ± 0.009) Å, using the relationship $a = \frac{1}{4}\sqrt{((3d_{111})^2 - c^2)}$ (where $d_{111}$ is the inter-planar lattice spacing determined from the (111) peak position). Thus, the film grows with an IP lattice expansion $\varepsilon_{||}$ = 0.74 %, and a corresponding OP lattice contraction $\varepsilon_{\perp}$ = -1.14 %, in agreement with the literature [36]. Due to a large lattice mismatch (≈ 10 %) with MgO (001) (*a* = 4.212 Å), the $Mn_3Ir$ (001) film couples only weakly to the substrate which, whilst sufficient to seed cube-on-cube growth, will not introduce epitaxial strain. Instead, $\varepsilon_{||}$ and $\varepsilon_{\perp}$ can be understood in terms of the film undergoing a small elastic distortion, where unit cell volume remains almost unchanged with respect to the bulk. We note that mosaicity is increased and lateral grain size suppressed with respect to (111) orientated films described subsequently. This can be explained by the simultaneous weak substrate-film interaction, combined with the inherent energetic instability of the (001) surface in fcc crystal structures, leading to frequent relaxation of the slight tetragonal distortion in $Mn_3Ir$, creating a higher areal density of grain boundaries and an enhanced rotation between neighboring grains.

Moving on to the characterization of $Mn_3Ir$ thin films with a (111) orientation deposited on $Al_2O_3$ (0001) substrates (a = 4.759 Å); in this case γ-$Mn_3Ir$ grows with the same fcc structure, but with the (111) crystal planes lying in the film plane in registry with the hexagonal substrate. Fig. 5(a) shows specular OP 2Θ-ω XRD patterns for 10 nm $Mn_3Ir$ films grown according to the previously discussed conditions, but now utilizing various buffer layers. In the case where $Mn_3Ir$ is grown directly on $Al_2O_3$ (0001), no crystalline structure is observed. This can be explained by a significant variation in interface free energies between the $Al_2O_3$ (0001) and $Mn_3Ir$ (111) surfaces making this growth mode unfavorable [40], a difference that may be reduced by the introduction of a buffer layer.

Therefore, two different buffer layers were employed; either 5 nm Pt or 5 nm TaN were deposited on $Al_2O_3$ substrates held at 500 °C. TaN was prepared according to the conditions described above, whilst Pt was deposited using a DC sputtering power of 50 W at a rate of 1.0 Å $s^{-1}$. Intense (111) and (222) peaks arising from both Pt and TaN are observed in Fig. 5(a), indicating that both films grow epitaxially on the hexagonal substrate with a sharp (111) texture [41]. Furthermore, both buffer layers seed a (111) orientation into the subsequently deposited 10 nm $Mn_3Ir$. A TaN buffer was chosen for further samples, giving the advantages of chemical stability, a smooth surface (with RMS roughness of < 3 Å confirmed by AFM), and a high resistance (measured as > 2 mΩ·cm via a four-probe method, in agreement with literature values [42]). From the OP 2Θ-ω XRD pattern in Fig. 5(a), a lattice parameter for TaN of (4.397 ± 0.004) Å is measured, which is close to the value for relaxed TaN thin films of 4.383 Å [42].

A lattice parameter value of (3.797 ± 0.001) Å is deduced for $Mn_3Ir$. This is very close to the bulk value, indicating that the film grows fully relaxed, in agreement with sputtered films prepared by Jara *et al.* [38]. OP grain size is calculated via the Scherrer formula to be (10.1 ± 0.3) nm, again demonstrating the correlation of grain size vertically with film thickness. A low mosaic spread in the film is measured as (0.478 ± 0.008) ° by recording a ω rocking curve XRD scan about the $Mn_3Ir$ (111) peak, displayed in Fig. 5(b). This low mosaicity, alongside the high-quality (111) crystal structure, can be attributed to the small lattice mismatch when using a TaN buffer layer, allowing relaxed film growth with minimal introduction of misfit dislocations or other defects. Indeed, between two periods of the $Al_2O_3$ substrate lattice and three of the TaN buffer, the lattice mismatch amounts to 2.0 %, which in turn reduces to just 0.8 % between the (111) oriented TaN and two periods of the Kagome planes of Mn atoms. The RMS roughness of the films is < 4 Å, as shown in Fig. 2(c), where the step and terrace topography of the $Al_2O_3$ substrate is observed via AFM through a 3 nm $Mn_3Ir$ (111) film.

Based on this analysis of lattice mismatch, the expected IP orientation of the stack is displayed in Fig. 6(a). This mode of epitaxial growth was confirmed by XRD pole figures, presented in Fig. 6(b), measuring the (002) reflections of (111) oriented TaN and $Mn_3Ir$ when the $[11\bar{2}0]$ axis of the substrate was aligned along $\phi = 0$ °. The sharp peaks observed evidence coherent IP crystallographic directions, whilst their six-fold symmetry suggests rotational twinning between (111) crystal planes [38,39]. The relative positions of the reflections confirm pseudo-hexagon-on-hexagon epitaxial growth throughout the stack, and allow the determination of the following epitaxial relationship, illustrated in Fig. 6(a): $Al_2O_3$ (0001) $[11\bar{2}0]$ $[\bar{1}100]$ || TaN (111) $[\bar{1}\bar{1}2]$ $[1\bar{1}0]$ || $Mn_3Ir$ (111) $[\bar{1}\bar{1}2]$ $[1\bar{1}0]$.

Additional TEM measurements on the (111) oriented samples investigated the film structure at the nanoscale. A cross section HAADF-STEM image of a 10 nm $Mn_3Ir$ film is displayed in Fig. 7(a), viewed along the $[1\bar{1}0]$ zone axis. The epitaxial growth of the TaN (111) buffer and $Mn_3Ir$ (111) film is clearly seen, demonstrating high-quality crystal structure with sharp interfaces and few defects. An absence of grain boundaries observed within the field of view suggests growth of large grains in the lateral direction, with a size of > 20 nm.

Fig. 7(b) shows a fast Fourier transform diffractogram of the lattice plane image in Fig. 7(a), with (*hkl*) diffraction peaks indexed. The positions of the diffraction spots confirm the epitaxial relationship between the layers. They further allow the determination of the predominantly IP lattice parameter along the [001] direction: (4.35 ± 0.09) Å for TaN and (3.84 ± 0.08) Å for $Mn_3Ir$. These agree, within uncertainty, with the OP lattice parameters measured from XRD, confirming the relaxed growth of $Mn_3Ir$ (111).

Finally, in order to study exchange bias induced by the AF $Mn_3Ir$, replicas of the above samples were prepared incorporating a ferromagnetic (FM) layer of 5 nm $Ni_{80}Fe_{20}$ (= Py, permalloy), grown from a $Ni_{80}Fe_{20}$ alloy target at a rate of 1.2 Å $s^{-1}$ via 75 W DC magnetron sputtering after samples had cooled to room temperature (RT), resulting in the growth of polycrystalline Py with composition $Ni_{(0.80\pm0.01)}Fe_{(0.20\pm0.01)}$ (measured by RBS).

**IV - Exchange bias**

Exchange bias (EB) was studied in bilayer samples of $Mn_3Ir$ / Py. EB occurs in coupled AF/FM systems, introducing a unidirectional anisotropy to the bilayer. This manifests itself as a shift in the FM magnetization hysteresis loop along the applied field axis, the exchange bias field ($\mu_0 H_{EB}$), as well as an enhancement of coercive field ($\mu_0 H_C$) [43,44]. EB is generally regarded as resulting from uncompensated spins at the interface of the AF, which exchange couple to moments in the FM layer [45,46]. These uncompensated AF spins are, in turn, strongly pinned in the direction of unidirectional anisotropy by AF domains that extend into the bulk of the film [47,48]. EB is set in a given direction at sufficient temperatures to overcome an energy barrier to AF domain reorientation, namely the blocking temperature, $T_B$. Here the application of an external magnetic field that saturates the FM will also align the coupled uncompensated moments, in turn leading to coherent orientation of bulk AF domains [49,50]. As the heterostructure is cooled, the preferential AF domain alignment becomes fixed below $T_B$ and exchange anisotropy is set in the direction of the external field [47,48]. The various characteristics of EB are determined by the thermal stability of the resulting AF domain walls, and hence depend intimately on film microstructure [51,52].

For the case of 10nm $Mn_3Ir$ / Py bilayers, $T_B$ lies above 300 K, meaning an EB can be stabilized at RT [24]. Fig. 8 (a) shows magnetization ($M$) measured as a function of IP field ($\mu_0 H$) for a 10 nm $Mn_3Ir$ (001) / Py bilayer; both as-deposited and after 30 min magnetic field annealing (FA) at 550 K and subsequent cooling in a 1 Tesla magnetic field ($\mu_0 H_{FA}$) applied along the [100] crystal direction (performed ex-situ in a furnace at a pressure < 9 × $10^{-6}$ mbar). The field annealing procedure did not result in modification of the crystal structure of the bilayer, as confirmed by XRD measurements. In the as-deposited state, no shift in the magnetization hysteresis ($MH$) loop can be seen. Following the IP field annealing procedure, a shift in the $MH$ loop of $\mu_0 H_{EB}$ = -28 mT, measured with applied field along the [100] axis in $Mn_3Ir$, demonstrates the onset of EB. This value of $\mu_0 H_{EB}$ is equivalent to a unidirectional anisotropy energy density (defined as $J_{\mathrm{K}} = M_{\mathrm{S}} d_{\mathrm{F}} \mu_0 H_{\mathrm{EB}}$, where $M_S$ is the saturation magnetization and $d_F$ is the thickness of the FM layer) of $J_K$ = 10.7 nJ $cm^{-2}$. Magnetization measured with external field along the perpendicular [010] crystallographic direction shows a hard axis response, confirming the unidirectional nature of the induced anisotropy. The negative shift of the hysteresis loops indicates the exchange anisotropy is set in the same direction as the field applied during annealing, because of the parallel coupling of the Ni magnetization in Py to interfacial Mn moments [53]. These uncompensated AF moments become, in turn, strongly pinned in their preferred direction by the dominant AF domain state in the bulk of the $Mn_3Ir$ film as the sample is cooled through $T_B$ [54].

Fig. 8(b) shows *MH* loops measured for a 10 nm $Mn_3Ir$ (111) / Py heterostructure with field applied along the $[\bar{1}\bar{1}2]$ crystalline direction after 30 min ex-situ 1 T IP field annealing at different temperatures, $T_{Anneal}$. In all cases, a negative shift of the hysteresis loop indicates the introduction of a unidirectional exchange anisotropy. The inset of Fig. 8(b) shows the variation in $\mu_0 H_{EB}$ with $T_{Anneal}$. A maximum $\mu_0 H_{EB}$ = -29 mT is achieved after IP field annealing at 550 K, corresponding to a unidirectional anisotropy energy density of $J_K$ = 11.5 nJ cm$^{-2}$ (in turn equivalent to $J_K \approx$ 0.1 erg cm$^{-2}$). Higher annealing temperatures lead to a degradation of $\mu_0 H_{EB}$, indicating that $T_B$ of these bilayers is close to 550 K, comparable to other values for epitaxially grown $Mn_3Ir$ films in the literature [35]. It is found that $\mu_0 H_{EB}$ is similar for both $Mn_3Ir$ orientations, contrary to Ref. [39] where larger $\mu_0 H_{EB}$ is measured for (111) textured films of similar thickness. This may be explained in our case by the (001) oriented samples containing a higher density of grain boundaries and larger mosaicity compared with the (111) films, as discussed above, which may act to enhance EB by introducing pinning sites to stabilize AF domain formation [55]. Furthermore, for both orientations we measure lower values of $\mu_0 H_{EB}$ compared to optimized values reported in the literature [56]. This is most likely because larger values of $\mu_0 H_{EB}$ are obtained in textured polycrystalline films containing much smaller grains [25] and a fraction of $L1_2$ ordered $Mn_3Ir$ phase [57].

On the other hand, the $T_B$ of 3 nm $Mn_3Ir$ / Py bilayers will lie below RT; it has been shown that $T_B$ decreases rapidly when $Mn_3Ir$ thicknesses is reduced below 5 nm [24,55] due to the reduced thermal stability of the AF domain state [58]. Fig. 9(a) shows the value of $\mu_0 H_{EB}$ measured for such a bilayer with (001) orientation after 1 T IP field cooling (FC) from 400 K to different temperatures, *T*. The inset to Fig. 9(a) shows an example of the individual *MH* loops measured at 5 K after zero field cooling (ZFC), +1 T IP field cooling and -1 T IP field cooling. Shifting of the *MH* loop along the applied field axis after field cooling, as opposed to zero field cooling, indicates the onset of EB at low temperatures and demonstrates the essential role of the external field ($\mu_0 H_{FC}$) in selecting a preferred direction for interfacial AF spins. The reversal of the unidirectional anisotropy after -1 T field cooling confirms the parallel coupling of the uncompensated Mn and FM Ni moments, whilst also showing the ability to manipulate interfacial magnetic structure and AF domain orientation as a function of field cooling.

Fig. 9(a) also shows the change in $\mu_0 H_{EB}$ when exchange anisotropy is induced by cooling the sample (and subsequently measuring) with magnetic field applied along different crystallographic directions. With EB along the [110] crystal axis, a higher blocking temperature ($T_B \approx 150$ K) is observed compared with the [100] axis ($T_B \approx 60$ K), as well as larger values of $\mu_0 H_{EB}$ at equivalent temperatures, in agreement with Ref. [35]. There is no obvious relation between microstructure (e.g. film terrace orientation measured by AFM) and this preferential axis for unidirectional anisotropy. However, we speculate that the enhanced stability of EB along the [110] crystalline direction is connected to the alignment of Mn moments in the γ-$Mn_3Ir$ structure at 45 ° to the cubic crystal axes, because the AF ordering of epitaxial $Mn_3Ir$ films has been shown to have a large influence on EB properties [36].

The measurement of $\mu_0 H_{EB}$ in a 3 nm $Mn_3Ir$ (111) / Py bilayer at different temperatures after 1 T IP field cooling from 400 K is shown in Fig. 9(b). In this case, no difference is seen in $\mu_0 H_{EB}$ with cooling field applied along different crystal directions. This may be due to six-fold IP crystalline symmetry in these samples, such that no direction provides a preferential axis for EB setting. Observed $T_B \approx 40$ K is found to be lower than (001) oriented $Mn_3Ir$, as is $\mu_0 H_{EB}$ at equivalent temperatures, with maximum $\mu_0 H_{EB} = -95$ mT at 5 K. Again, this may be attributable to higher quality epitaxial growth of (111) films, introducing less defects and grain boundaries to stabilize AF domains at a given temperature [55]. The inset of Fig. 9(b) shows individual *MH* loops recorded at 5 K following IP field cooling with different external field strengths; $\mu_0 H_{EB}$ is invariant with field strength as expected, again indicating the potential uses of EB in manipulating AF order using low applied fields. This degree of control, as well as its close relation to domain structure, means exchange bias may play a valuable role when utilizing chiral AFs for spintronic applications.

Under such circumstances, an important consideration is the EB training effect. Here the measured $\mu_0 H_{EB}$ and $\mu_0 H_C$ decrease over the course of successive external applied magnetic field cycles. This arises because a portion of the uncompensated AF moments at the interface, that contribute to exchange coupling after the initial field cooling procedure, are only weakly pinned to the bulk AF structure. They are thus free to follow the reversing magnetization of the FM layer, and so are re-orientated by the external field [59]. In Fig. 10(a) the change in $\mu_0 H_{EB}$ and $\mu_0 H_C$ measured over the course of consecutive training field cycles is shown for a (001) oriented 3 nm $Mn_3Ir$ / Py bilayer, following 1T IP field cooling from 300 K to different temperatures. In all cases, it is observed that, after at most 4 applied field cycles, both $\mu_0 H_{EB}$ and $\mu_0 H_C$ reach equilibrium values and do not change further. At this point all weakly pinned uncompensated Mn spins are relaxed. The remaining exchange bias is modulated by interfacial AF spins that are strongly coupled to the bulk $Mn_3Ir$ domain state [46]. Similar results are seen for 3 nm $Mn_3Ir$ (111) / Py heterostructures (not shown).

These resulting values of exchange bias therefore depend on the stability of the AF order, and hence on the temperature to which the bilayer was field cooled. Whilst before training both $Mn_3Ir$ orientations show large $\mu_0 H_{EB} \geq -95$ mT at 5 K, the maximum post–training $\mu_0 H_{EB} = -77$ mT for a 3nm $Mn_3Ir$ (001) / Py bilayer and $\mu_0 H_{EB} = -27$ mT for the (111) orientation. The dramatic decrease in $\mu_0 H_{EB}$ for the 3nm $Mn_3Ir$ (111) / Py bilayer may indicate the significant contribution to the initial exchange bias setting of weakly coupled uncompensated Mn moments in ultrathin films of this orientation, as discussed further in our subsequent work [60].

Finally, in order to confirm the $T_B$ of bilayers with ultrathin $Mn_3Ir$, further temperature dependent measurements of exchange bias were performed. In Fig. 10(b) the variation in $\mu_0 H_{EB}$ and $\mu_0 H_C$, extracted from magnetization hysteresis loops measured at 5 K after 1 T IP field cooling from different starting temperatures, $T_{Start}$, is shown for a 3 nm $Mn_3Ir$ / Py bilayer with (111) orientation. A sharp decrease in both $\mu_0 H_{EB}$ and $\mu_0 H_C$ is observed when cooling from temperatures below 40 K, indicating that $T_{Start}$ is no longer completely above the maximum of the bilayer's $T_B$ distribution, and thus insufficiently energetic to fully reorient AF domains in order to obtain maximum exchange bias. Variation in grain size within the $Mn_3Ir$ film results in a distribution of these activation energies and hence of $T_B$, accounting for the steady decrease in $\mu_0 H_{EB}$ towards zero as $T_{Start}$ is further decreased [25].

## V – Conclusion

In summary, recipes for the deposition of γ-$Mn_3Ir$ with (001) orientation on MgO substrates, and with (111) orientation on TaN buffered $Al_2O_3$ substrates, are reported. A combination of XRD and TEM analysis demonstrates the epitaxial growth of the thin films and the resulting high-quality crystal structure, with $Mn_3Ir$ (111) films in particular showing low mosaicity and large grain size. EB was studied in bilayer samples, with values up to $\mu_0 H_{EB}$ = -29 mT ($J_K$ = 11.5 nJ $cm^{-2}$) achieved after 1 Tesla in-plane field annealing at 550 K. For heterostructures with ultrathin antiferromagnet layers, exchange bias is observed below room temperature, with $T_B$ ≈ 40 K in 3 nm $Mn_3Ir$ (111) / Py samples and a notable dependence of exchange coupling on in-plane crystalline direction in 3 nm $Mn_3Ir$ (001) / Py bilayers. Here a higher $T_B$ ≈ 150 K and larger values of $\mu_0 H_{EB}$ are measured when unidirectional anisotropy is set along the [110] crystallographic axis. These findings may inform future studies of spin-orbit torques in such heterostructures, whilst our present results provide a springboard for further investigation of epitaxial thin films of $Mn_3Ir$ and other non-collinear antiferromagnets. In particular, we highlight the subtle influence of crystal quality and how it may be manipulated by epitaxial engineering, an important consideration when utilizing this class of materials in topological *chiralitronic* devices.

## Acknowledgments

We thank Dr Andrew Kellock for RBS measurements, and are grateful to Dr Holger L. Meyerheim for assistance with XRD analysis. This work was partially funded by ASPIN (EU H2020 FET open grant 766566).

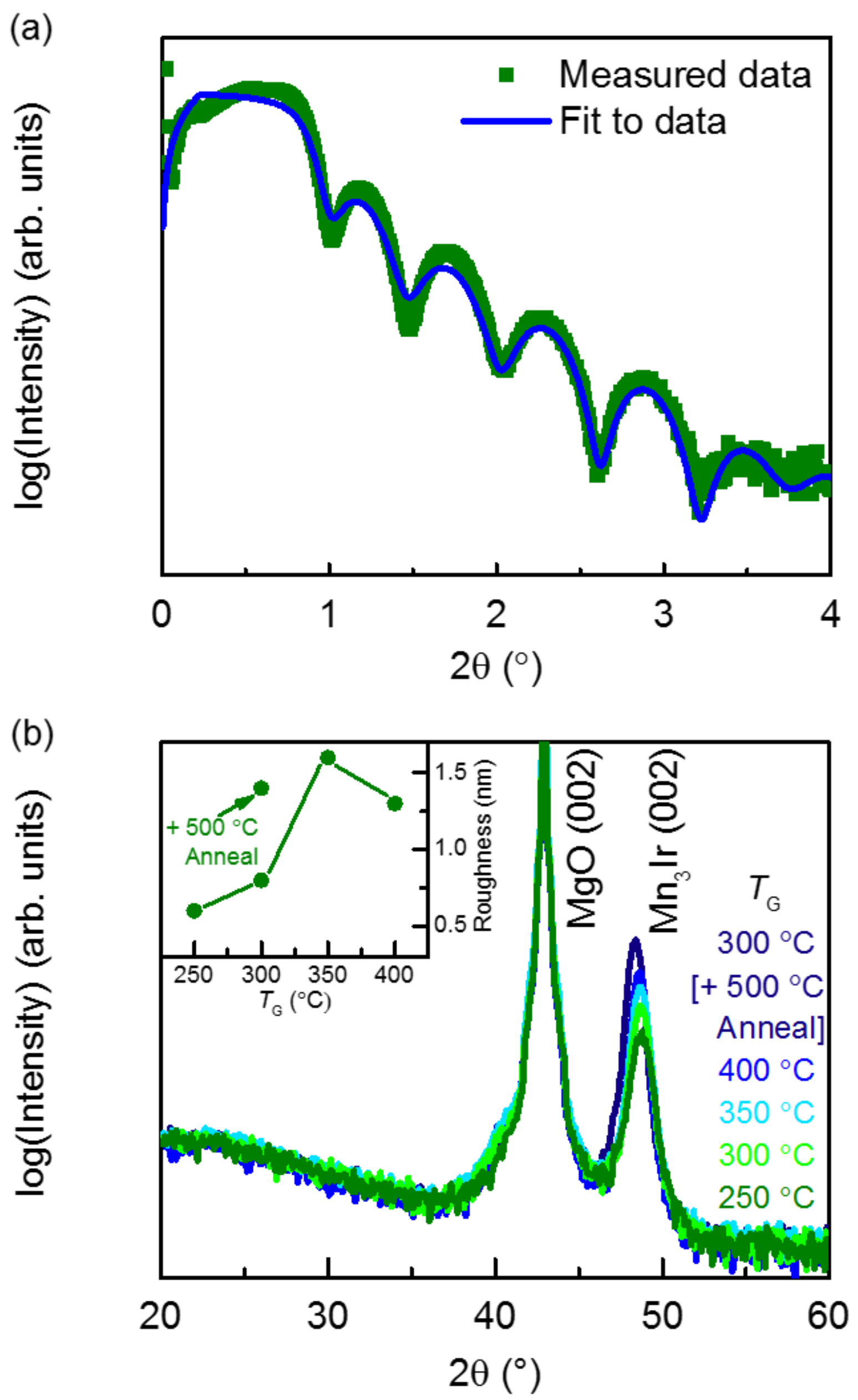

FIG. 1. (a) Measured X-ray reflectivity data from a 10 nm $Mn_3Ir$ (001) film, with fit to determine layer thicknesses. (b) XRD 2Θ-ω scans measured for 10 nm $Mn_3Ir$ (001) films grown at different temperatures (inset shows RMS roughness measured by AFM for each of these samples).

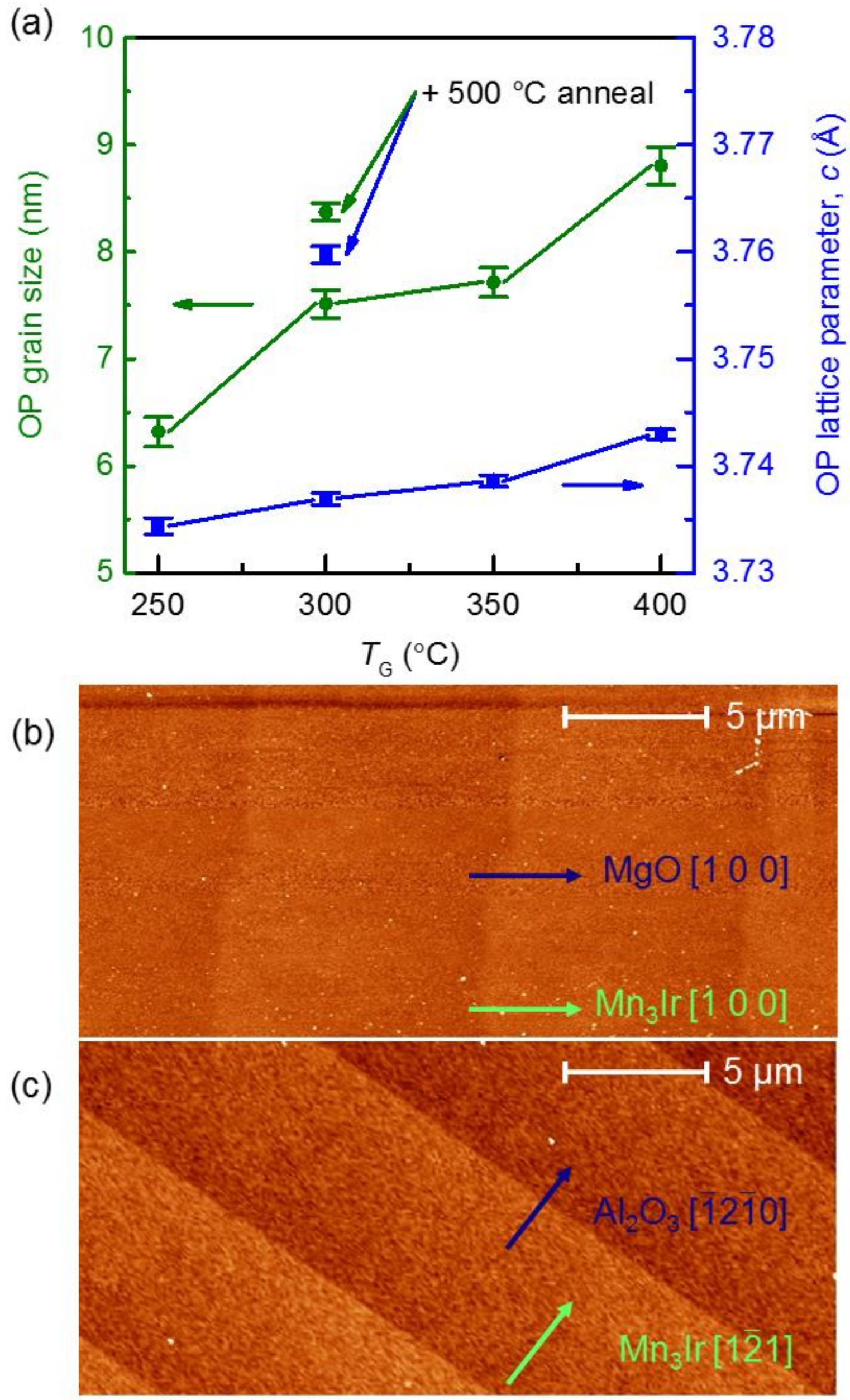

FIG. 2. (a) Dependence of grain size and OP lattice parameter on growth temperature. AFM topography maps of 3 nm $Mn_3Ir$ films with (b) (001) and (c) (111) orientation.

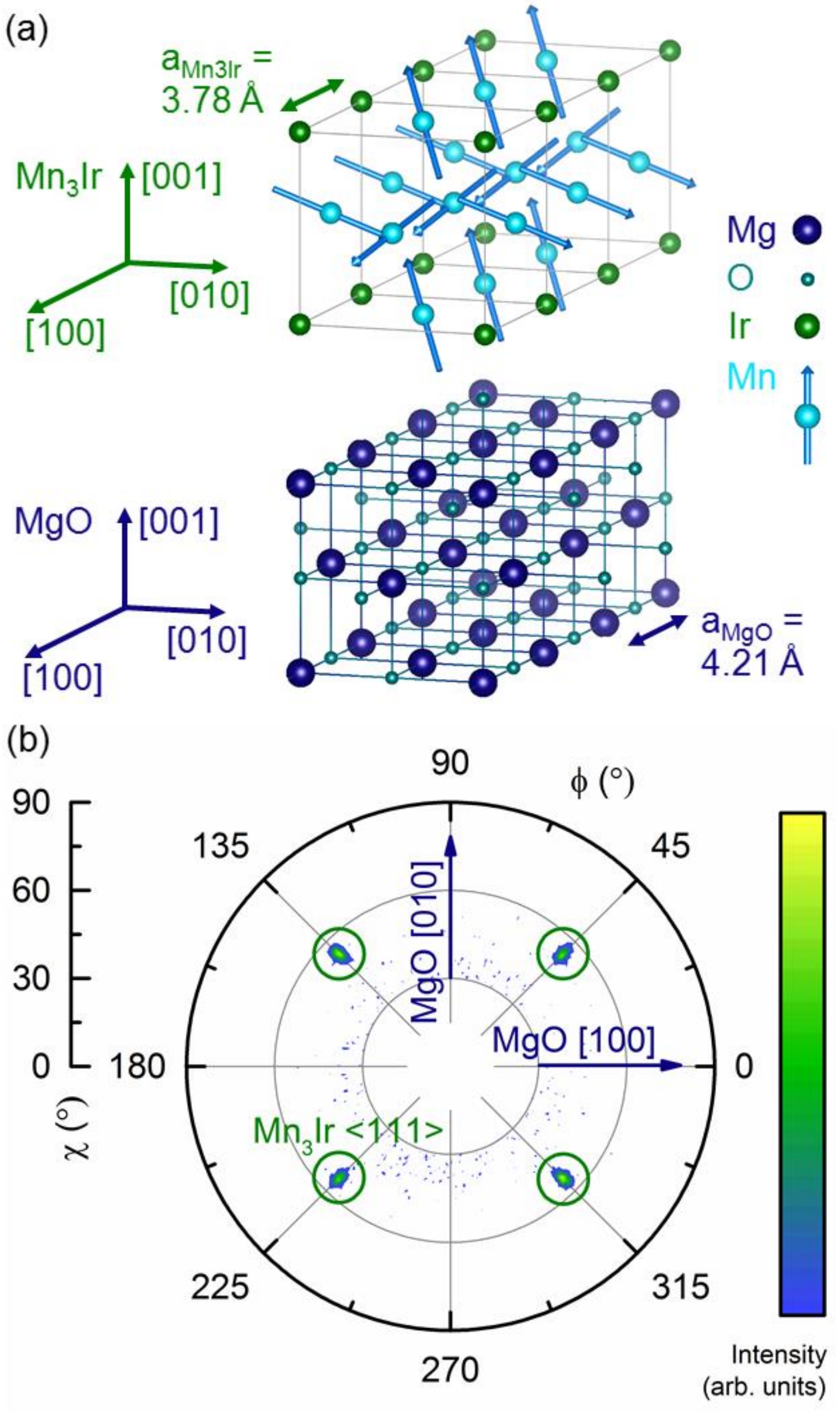

FIG. 3. (a) Crystal and magnetic structure of $L1_2$ ordered $Mn_3Ir$ with [001] axis directed out-of-plane, demonstrating cube-on-cube epitaxy with a (001) oriented MgO substrate. (b) XRD χ-ϕ pole figure measuring <111> peaks in a 10 nm $Mn_3Ir$ film with (001) orientation, aligned such that the [100] and [010] axes of the MgO substrate are directed along ϕ = 0 ° and ϕ = 90 ° respectively.

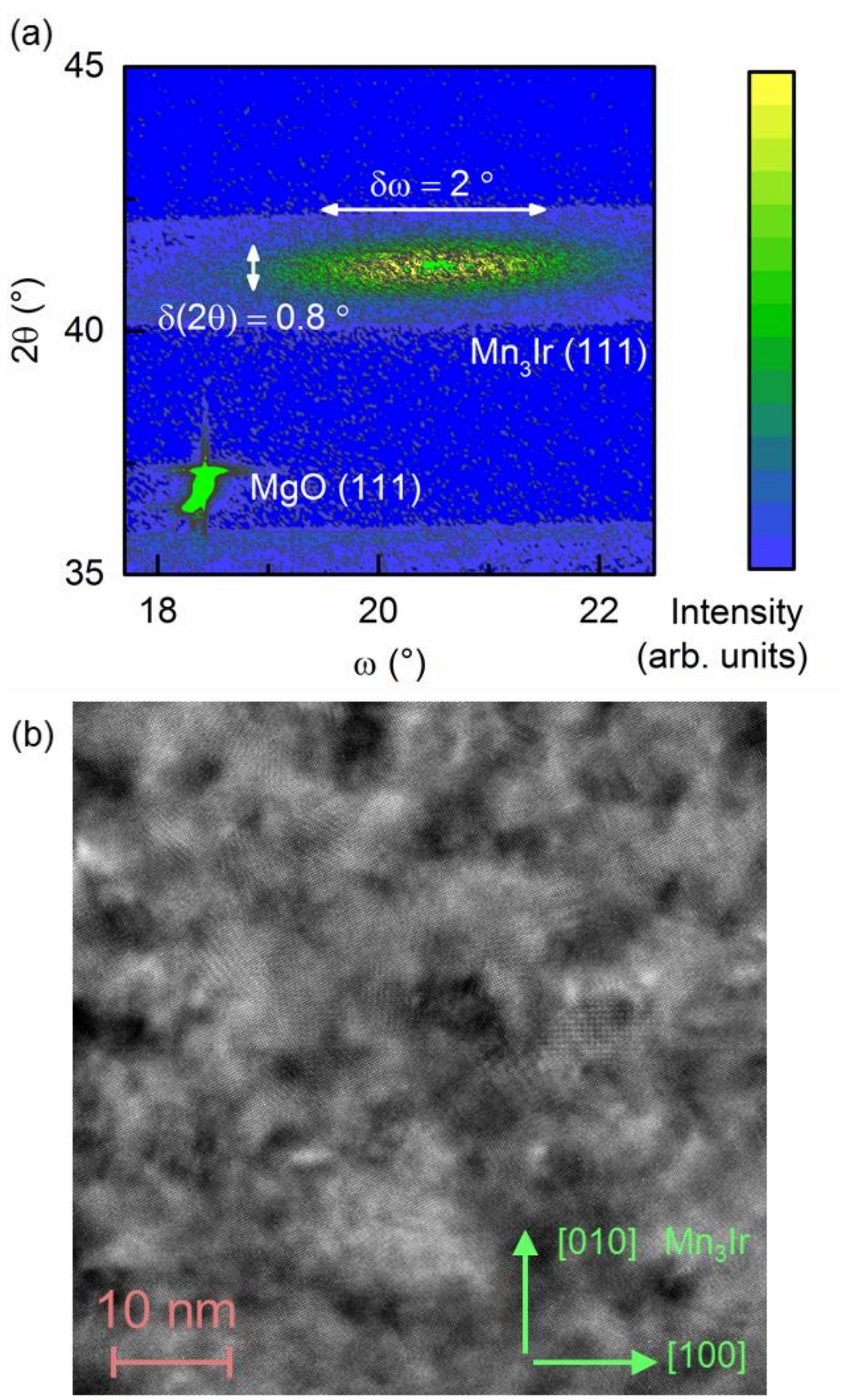

FIG. 4. (a) XRD off-specular scan map measuring (111) reflections from a 10 nm $Mn_3Ir$ film with (001) orientation and the MgO (001) substrate upon which it is grown. (b) Plane-view TEM image of a 3 nm $Mn_3Ir$ (001) thin film film along the [001] *c*-axis, prepared using ion-beam backside thinning.

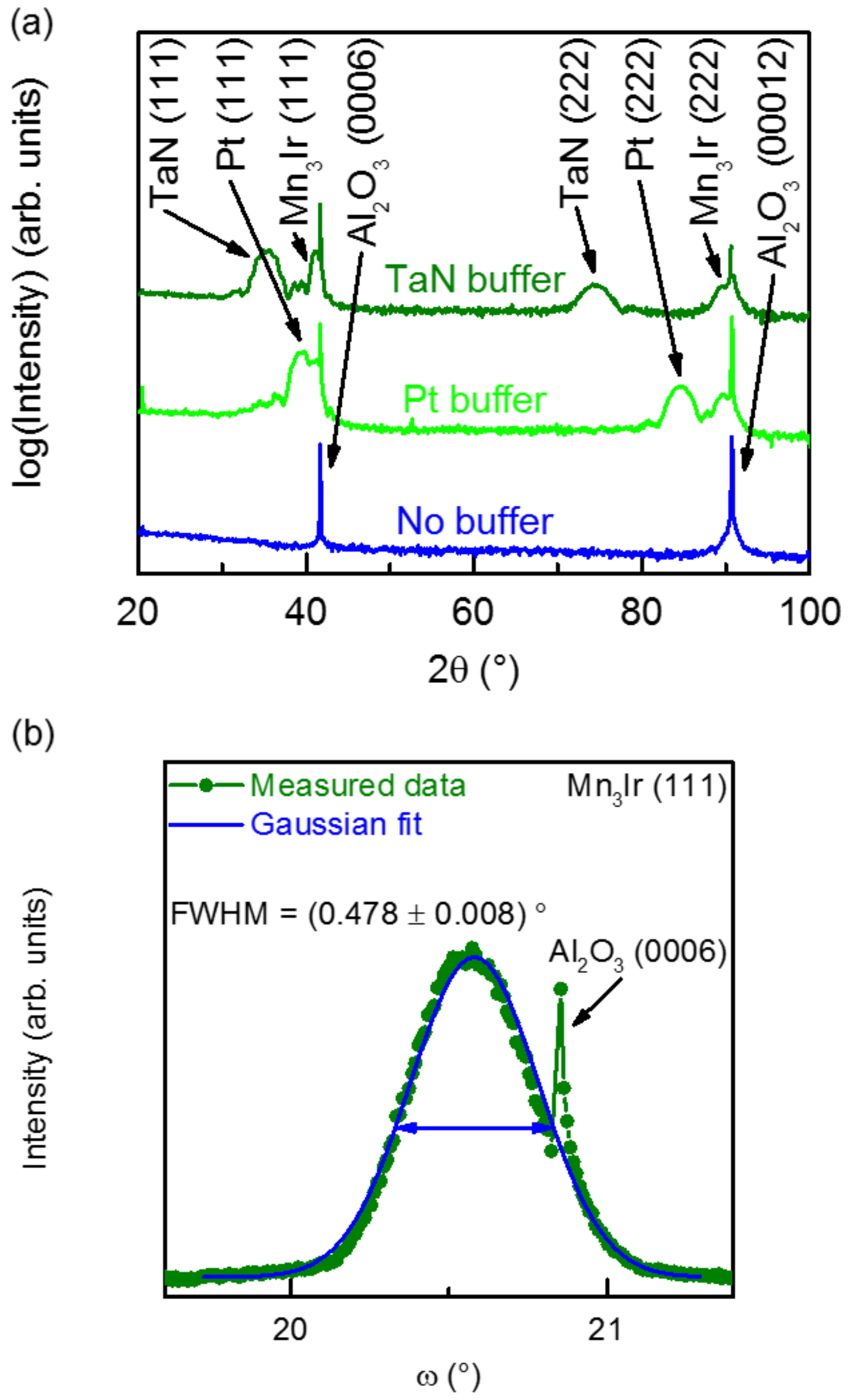

FIG. 5. (a) XRD 2Θ-ω patterns measured for 10 nm $Mn_3Ir$ (111) films grown using different buffer layers (scans are offset for clarity). (b) XRD ω rocking curve for a 10 nm $Mn_3Ir$ (111) film grown on a TaN buffer layer, with fit to determine FWHM.

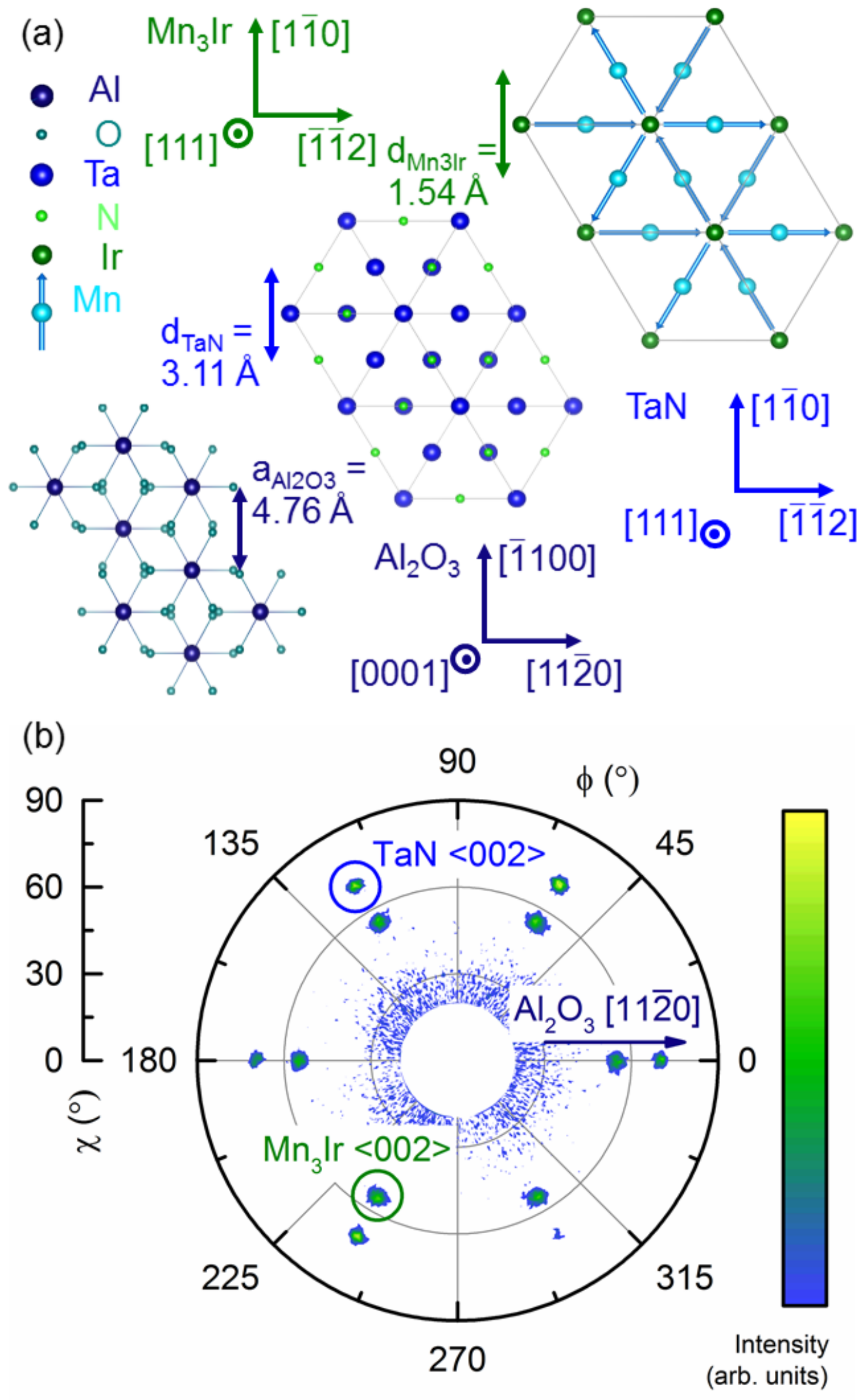

FIG. 6. (a) Crystal and magnetic structure of (111) planes in *L*$1_2$ ordered $Mn_3Ir$, of (111) planes in TaN and of (0001) planes in $Al_2O_3$, showing the epitaxial relation between them as viewed along the OP axis. (b) XRD χ-ϕ pole figure measuring <002> peaks in a 10 nm $Mn_3Ir$ film with (111) orientation and a 5 nm TaN (111) buffer layer, aligned such that the $[11\bar{2}0]$ axis of the $Al_2O_3$ substrate is directed along ϕ = 0 °.

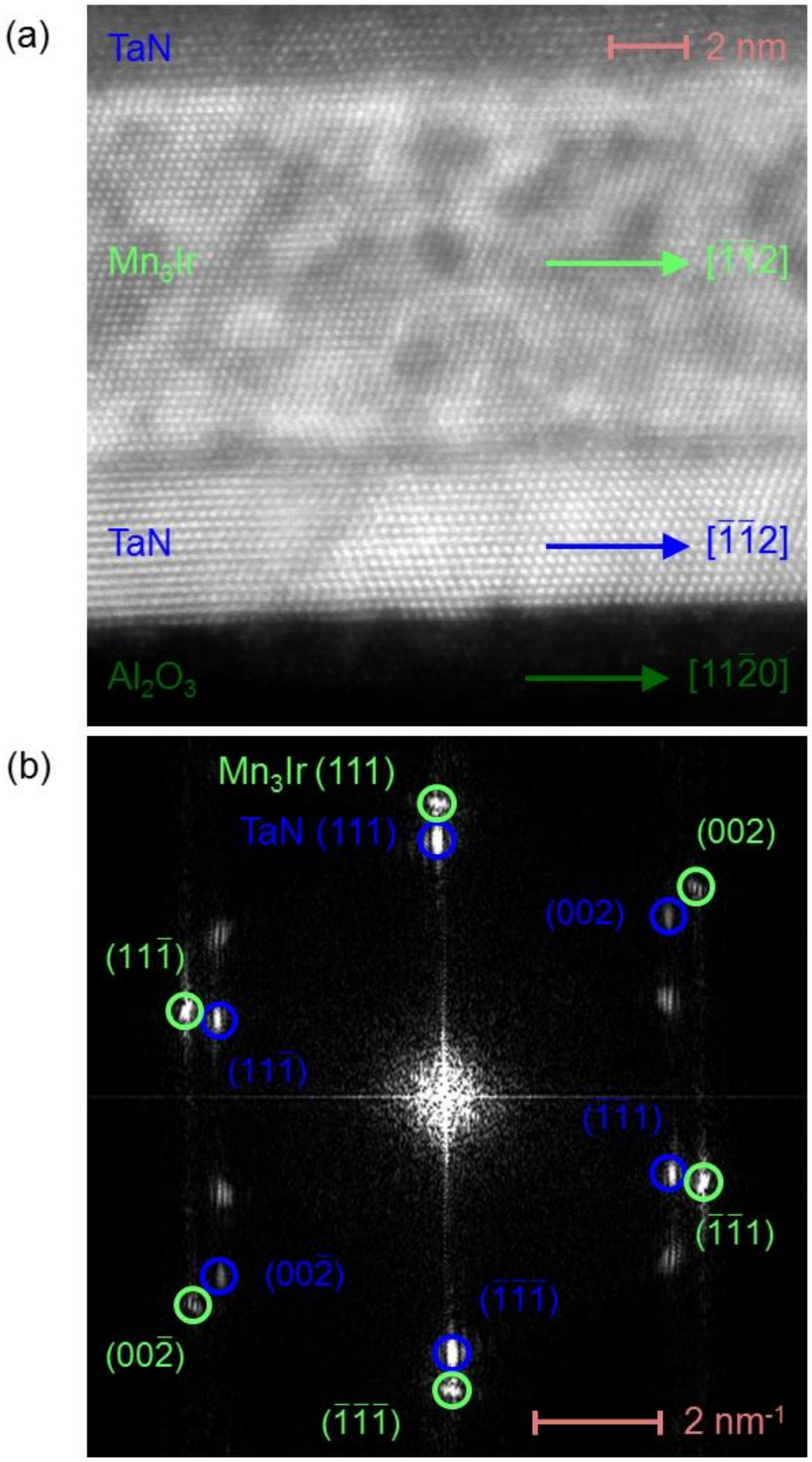

FIG. 7. (a) Cross sectional HAADF-STEM image of a 10 nm $Mn_3Ir$ (111) film, grown on an $Al_2O_3$ substrate with (0001) orientation using a 5 nm (111) textured TaN buffer layer, viewed along the $[1\bar{1}0]$ zone axis. (b) Diffractogram (Fast Fourier Transform) of the above experimental image. The epitaxial correlation between the $Mn_3Ir$ and TaN lattices is demonstrated by the corresponding indexed reflections.

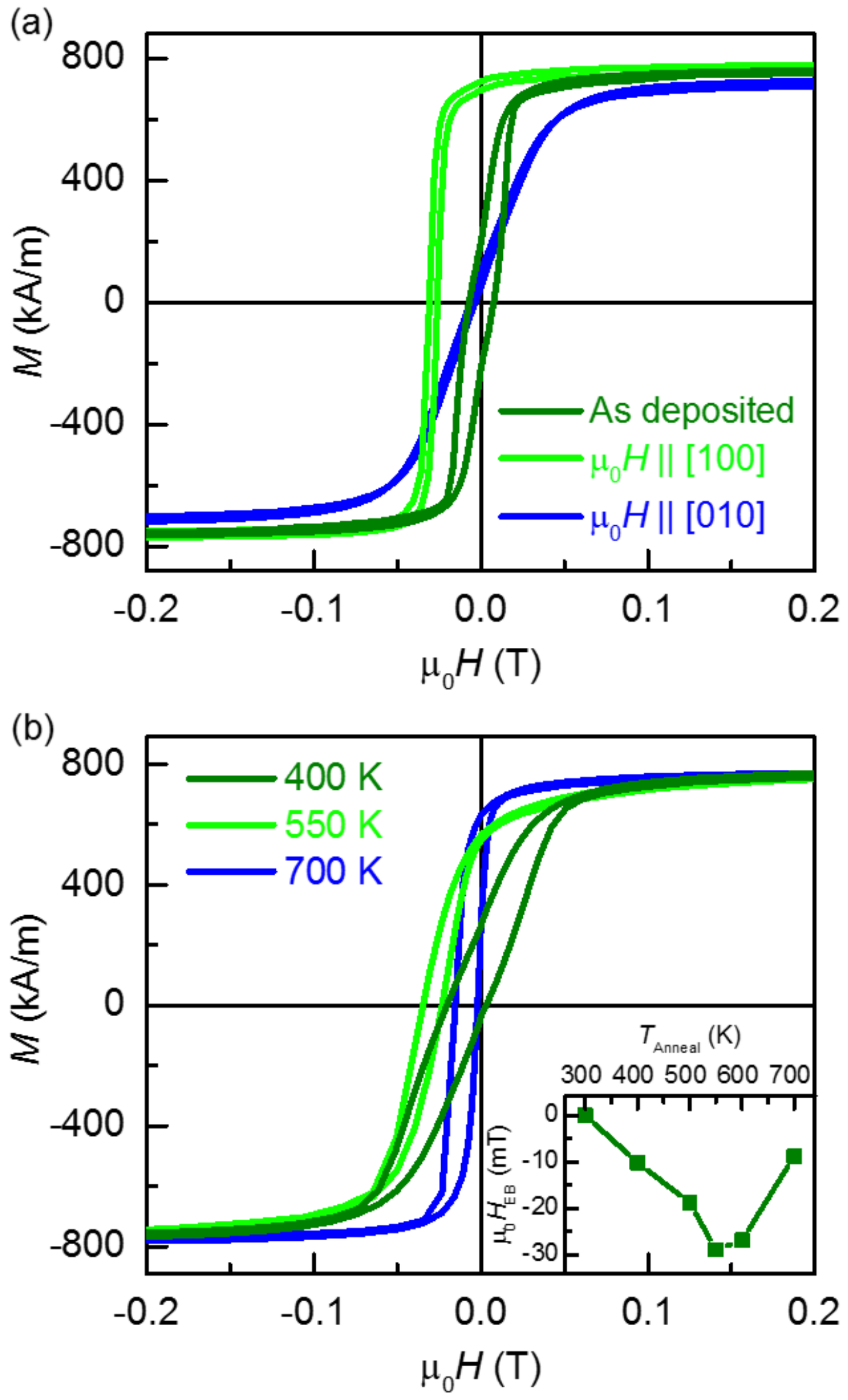

FIG. 8. (a) Magnetization hysteresis loops at 300 K for an as-deposited 10 nm $Mn_3Ir$ (001) / Py bilayer, and for the same sample after 1 T IP field annealing at 550 K (with $\mu_0 H_{FA}$ || [100]), with IP measurement field directed parallel and perpendicular to $\mu_0 H_{FA}$. (b) Magnetization hysteresis loops at 300 K for a 10 nm $Mn_3Ir$ (111) / Py bilayer after 1 T IP field annealing at different temperatures (with $\mu_0 H_{FA}$ || $[\bar{1}\bar{1}2]$) (inset shows variation of $\mu_0 H_{EB}$ with annealing temperature).

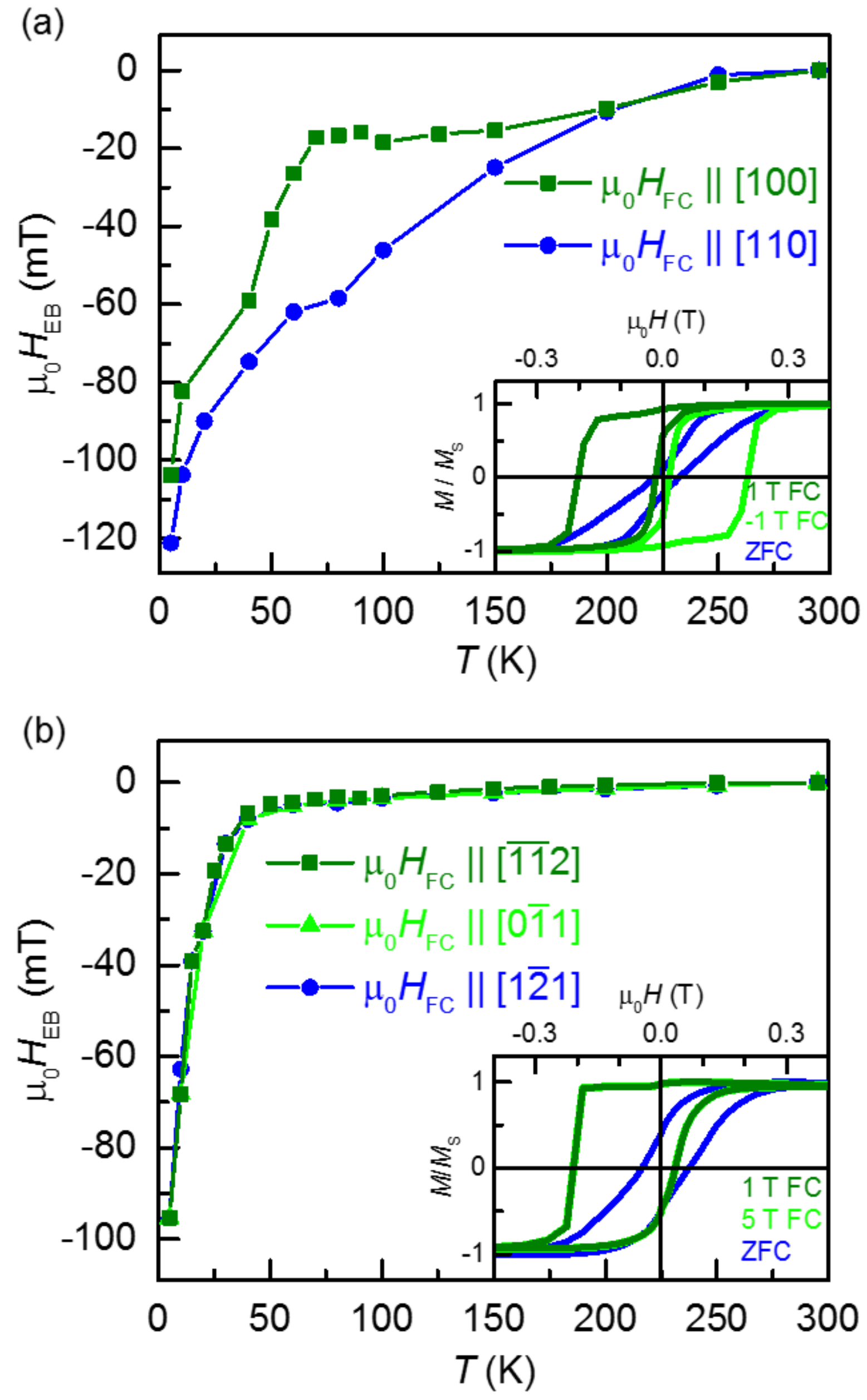

FIG. 9. (a) $\mu_0 H_{EB}$ measured at different temperatures after 1 T IP field cooling from 400 K for a 3 nm $Mn_3Ir$ (001) / Py bilayer, with $\mu_0 H_{FC}$ || [100] and [110] crystal axes (inset shows normalized magnetization hysteresis loops recorded at 5 K after different field cooling protocols with $\mu_0 H_{FC}$ || [100]). (b) $\mu_0 H_{EB}$ measured at different temperatures after 1 T IP field cooling from 400 K for a 3 nm $Mn_3Ir$ (111) / Py bilayer, with $\mu_0 H_{FC}$ || $[\bar{1}\bar{1}2]$, $[0\bar{1}1]$ and $[1\bar{2}1]$ crystal axes (inset shows normalized magnetization hysteresis loops recorded at 5 K after different field cooling protocols with $\mu_0 H_{FC}$ || $[\bar{1}\bar{1}2]$).

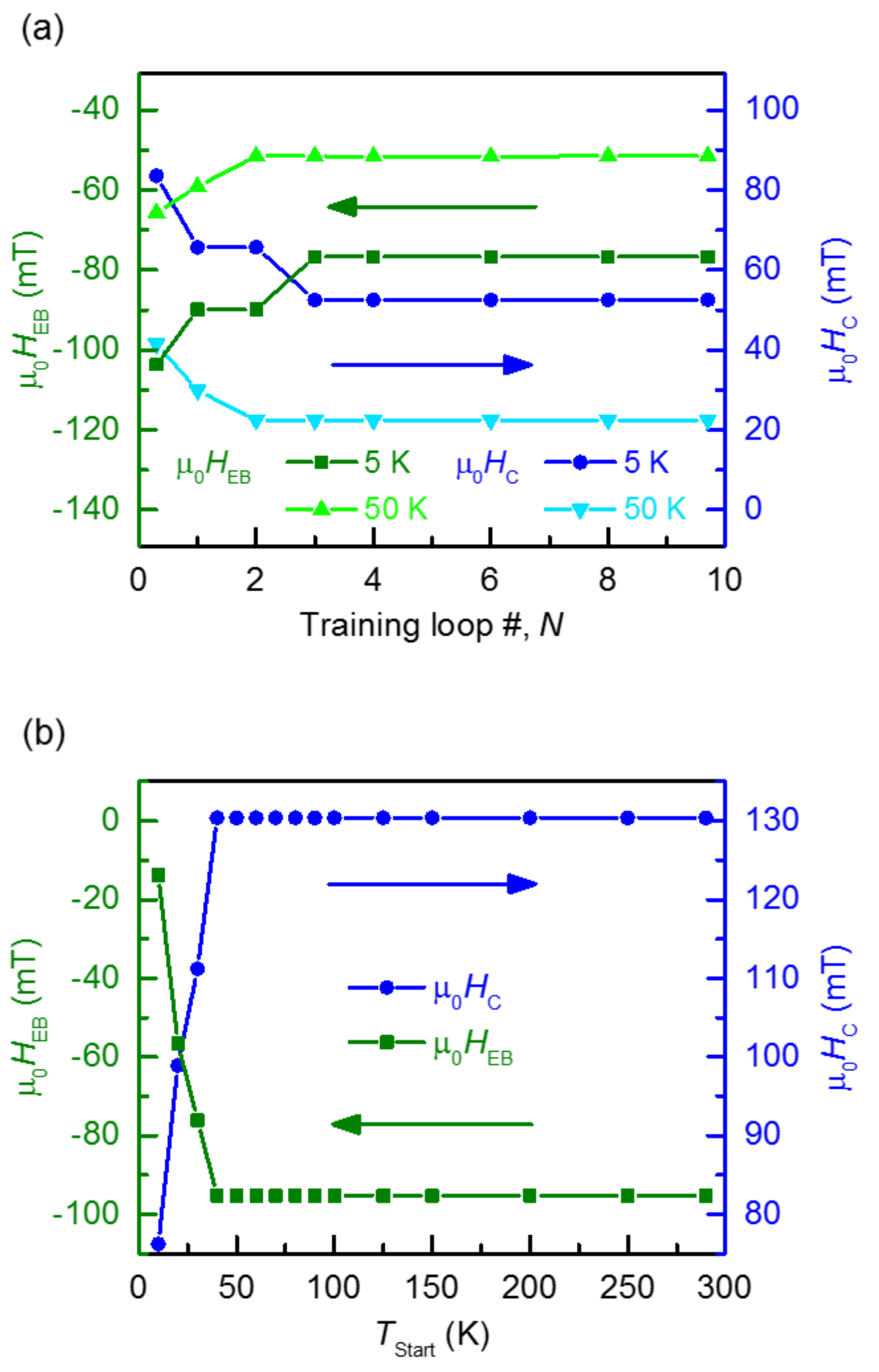

FIG. 10. (a) Exchange bias training effect showing variation of $\mu_0 H_{EB}$ and $\mu_0 H_C$ with successive measurement field cycles at different temperatures after 1 T IP field cooling from 300 K for a 3 nm $Mn_3Ir$ (001) / Py bilayer (with $\mu_0 H_{FC}$ || [100]). (b) Blocking temperature distribution showing $\mu_0 H_{EB}$ and $\mu_0 H_C$ measured at 5 K after 1 T IP FC from different starting temperatures for a 3 nm $Mn_3Ir$ (111) / Py bilayer (with $\mu_0 H_{FC}$ || [$\bar{1}\bar{1}$2]).